\documentstyle[aps,prb]{revtex}
\begin{document}
\draft

\input psfig
\twocolumn[\hsize\textwidth\columnwidth\hsize\csname @twocolumnfalse\endcsname

\title{Fermi Liquid Theory and Ferromagnetic Manganites
at Low Temperatures}
\author{M.\ O.\ Dzero$^{1}$, L. P. Gor'kov$^{1,2}$ and V.\ Z.\ Kresin$^{3}$}
\address{$^1$National High Magnetic Field Laboratory, Florida State
University, Tallahassee, FL 32310}
\address{$^2$L.D. Landau Institute for Theoretical Physics, Russian Academy
of Sciences, 117334 Moscow, Russia}
\address{$^3$Lawrence Berkeley Laboratory, University of California, Berkeley, 
CA 94720}
\date{\today}
\maketitle

\begin{abstract}

~Fermi ~liquid ~characteristics
~for ~ferromagnetic ~manganites, ~A$_{1-x}$B$_x$MnO$_3$,
are evaluated in the tight-binding approximation and compared with experimental
data for the best studied region $x\simeq0.3$.The bandwidths change only 
slightly for different compositions. The Sommerfeld coefficient, $\gamma$, the
$T^2$-term in resistivity and main scales in optical conductivity agree well
with the two band model. The ``2.5'' - transition due to a ``neck'' forming
at Fermi surface, is found at $x=0.3$. The mean free path may change from 3 to 80
interatomic distances in the materials, indicating that samples' quality remains
a pressing issue for the better understanding of manganites. 
\end{abstract}
\narrowtext
]

Experiments on doped ``pseudocubic'' manganites, A$_{1-x}$B$_x$MnO$_3$,
primarily engendered by prospects of applications of the ``colossal''
magnetoresistance (CMR) effect, unveiled unexpected richness of new
phenomena pressing for a fundamental interpretation (for review see, e.g.
\cite{one,two}). The consensus is that the ``double exchange'' (DE) mechanism
\cite{three}, together with the Jahn-Teller (JT) instability of the degenerate
$e_{2g}$-term at the Mn$^{3+}$-site pre-determine the
$(x,T)$-phase diagram of the low doped manganites. The
one-electron Hamiltonian of the form \cite{four}:
\begin{eqnarray}
\hat{H}=\sum_{i,\delta} \left ( \hat{t}_{i,i+\delta}-
J_H({\bf S}_i\cdot \hat{\bbox{\sigma}})+
g \hat{\bbox{\tau}} \cdot {\bf Q}_i+J_{def}{\bf Q}_i^2 \right )
\label{eq:a1}
\end{eqnarray}
accounts for the competition
between the first two terms in (\ref{eq:a1}) ($\hat{t}_{i,\delta}$ being a
tunneling matrix, while $J_H$ is the strong intrasite Hund's
coupling) which, in accordance with the DE-mechanism, tend to form
ferromagnetic electronic bands, and localizing
effects of the thermal disorder in the JT-degrees of freedom (the last
two terms).  The method \cite{four} interpolates the
high- and low-temperature regimes in ``low doped''
($x<0.5$) manganites in terms of crossover between ``small'' and
``large'' polarons conduction \cite{four,five,six} . 
It is not apt to draw conclusions 
regarding the ground state symmetry.

Another issue is whether
the electron-electron interactions are of principle  
importance for manganites.  The stoichiometric
compound, LaMnO$_3$, is an antiferromagnetic (the A-phase) insulator and
preserves the insulating low-T behavior even at doping, $x$, below
$x_{cr}\simeq0.16-0.17$. These features are often treated in the literature
in terms of a Mott-Hubbard state.
\indent
It has been demonstrated in  \cite{seven}  
that properties of LaMnO$_3$ may be easily
rationalized in terms of the band insulator provided static
JT-distortions are taken into account in the electron band structure.
It was also argued that the critical concentration
$x_{cr}\simeq0.16-0.17$ for the onset  of a metallic (low T)
conductivity  is precisely the percolation threshold.  (The
CMR-phenomenon at T$_c$ itself may be interpreted in the percolative
terms as well \cite{eight} ).  
Hence, at $x$ close enough to $x_{cr}$
doped manganites exist as a highly inhomogeneous ``mixture'' of tiny
islands of coexisting phases of a size, limited by the Coulomb forces
(typically, of the order of 10-20 $\AA$ \cite{nine,ten}).
Crossover from percolation regime to homogeneous ferromagnetic
phase seems to take place below $x \sim 0.3$ (see discussion below).
In what follows we analyze available low temperature 
experimental data for doped manganites with $x \sim 0.3$.

As $x$ increases above  $x_{cr}$, percolation, 
being a critical phenomenon, may cede
soon to the onset of the homogeneous ferromagnetic phase, when
screening becomes effective.  In
fact, some good samples of La$_{1-x}$Sr$_x$MnO$_3$ show low temperature
resistivity in the range $10^{-4}-10^{-5}(\Omega -cm)$  \cite{odin}.

In what follows we analyze available low temperature
experimental data for doped manganites from this point of view.  
Our conclusions are that the band model (or Fermi liquid)
approach gets valid at low temperatures, especially for
high quality samples. Data are hindered by sample's quality
with local inhomogeneities forming scattering centers.
For a number of compositions depending on tolerance factor,
conductivity is close to it's value in the mobility edge regime. 
Data on low frequency optical conductivity $\sigma(\omega)
(\omega < 1 eV)$ remain a controversial 
issue (at low T), although agree qualitatively with the
two band model  \cite{seven}.

{\it Summary of the Theoretical Results.} -- We adopt the model and
notations of ~\cite{seven}.  
The electron spectrum consists of the two branches:
\begin{eqnarray}
\varepsilon_{\pm}({\bf p})&=& -|A|\cdot\left(c_x+c_y+c_z\pm 
R({\bf p})\right),\label{eq:a2} \\ 
~R({\bf p})&=&\sqrt{c_x^2+c_y^2+c_z^2-c_xc_y-c_xc_z-c_yc_z}.
\label{eq:a2pr}
\end{eqnarray}
Here for brevity: $c_i=\cos \tilde{p_i}~,~\tilde{p_i}=p_ia$.

We calculate the concentration dependence of the Fermi-level,
$E(x)$, the density of states (DOS), $\nu(x)$, the spin stiffness,
$D(x)$, and the whole magnon spectrum, $\omega({\bf k},x)$, and the
conductivity, $\sigma(\omega, x)$, both the Drude and the optical
(interband) components.  Theoretical results depend only on the single hopping
integral, $|A|$.  In addition, for the stoichiometric LaMnO$_3$
the optical data for $\sigma(\omega)$ at $x=0$  \cite{pjat,shes}  allow 
to evaluate the electron coupling with the JT-distortions.

To find the spin waves spectrum, present 
deviations from the average spin, $\langle S_z\rangle$ for the localized
$t_{2g}$ - spins $({\bf s}={\bf S}-\langle S_z\rangle)$ as:
\begin{eqnarray}
{~s}^+({\bf q})=(2\langle S_z\rangle)^{1/2}&&\hat{b}({\bf q}),
{~s}^-({\bf q})=(2\langle S_z\rangle)^{1/2}\hat{b}^+({\bf q}),
\nonumber\\
&&{~s}_z({\bf q})=(\hat{b}^+\hat{b})_{\bf q},
\label{eq:a3}
\end{eqnarray}
($\hat{b}^+, \hat{b}$-the magnon's operators).  The first ($\delta
E_1$) and second ($\delta E_2$) order corrections to the
ground state are calculated as perturbations in:
\begin{equation}
\hat{V}=-J_H\sum_i{\bf s}_i(\hat{a}_i^+{\boldmath
\sigma}\hat{a}_i)=-J_H\sum_i{\bf s}_i{\bf n}_i.
\label{eq:a4}
\end{equation}

For $\delta E_2$, the matrix elements in (\ref{eq:a4}) are
of the form:
\begin{eqnarray}
V_{{\bf p}-{\bf k},\uparrow; 
{\bf p}\downarrow}^{l,l'}&=&-J_H\langle\uparrow
|s^{\pm}({\bf k})|\downarrow\rangle\nonumber\times\\
&&\times\left(\alpha_{{\bf p}-{\bf
k}}^l\alpha_{\bf p}^{\ast 
l'}+\beta_{{\bf p}-{\bf k}}^l\beta_{\bf p}^{\ast l'}\right),
\label{eq:a5}
\end{eqnarray}
(see (\ref{eq:a8}). 
As for $\delta E_1$, its only role is to secure the proper behavior of
the magnon spectrum at $k\rightarrow 0$).  
One obtains:
\begin{eqnarray}
\delta E_2=&&2 J_H^2\sum_{\bf k}\langle S_z\rangle
\hat{b}^+({\bf k}) \hat{b} ({\bf k})\times\nonumber\\
&&\times\sum_{l,{\bf p}}\left(
\sum_{l'}\frac{|\alpha_{\bf p}^l\alpha_{{\bf p}+{\bf k}}^{\ast
l'}+\beta_{\bf p}^l\beta_{{\bf p}+{\bf k}}^{\ast
l'}|^2}{E_{\uparrow}^l({\bf p})-E_{\downarrow}^{l'}({\bf p}+{\bf k})}\right),
\label{eq:a6}
\end{eqnarray}
where
\begin{eqnarray}
E_{\uparrow, \downarrow}^{l, l^{\prime}}({\bf p})&=&
\mp J_H\langle S_z\rangle
+\varepsilon_{l, l^{\prime}}({\bf p}).
\label{eq:a7}
\end{eqnarray}
Both sums in (\ref{eq:a6}) run over 
$l,l^{\prime}=\pm$ (Eq.(~\ref{eq:a2})).  
The summation over $l$ and ${\bf p}$ is limited by the occupied states
$(\uparrow)$ only.  The coefficients $(\alpha_{\bf p}^l, \beta_{\bf
p}^l)$ above are for the Bloch's states on the basis,used in \cite{seven}:
\begin{equation}
\alpha_{\bf p}^{l,l'}=\left(t_{12}/2|t_{12}|\right)^{1/2},
~\beta_{\bf p}^{l,l'}=\pm\left(t_{21}/2|t_{12}|\right)^{1/2},
\label{eq:a8}
\end{equation}
(here $t_{12}, t_{21}$ are the off-diagonal elements of the 
hopping matrix $\hat{t}$({\bf p}) 
in (\ref{eq:a1}) on this basis).

Assuming in (\ref{eq:a6}, \ref{eq:a7}) $J_H\gg
|A|$ ($J_H \langle S_z\rangle$ is of the order 
of 1.5 eV  \cite{one} , 
estimates below produce
for $|A| \sim 0.1$ eV),
expansion (\ref{eq:a6}) in $|A|/J_H$ 
gets equivalent to the series of the Heisenberg spin Hamiltonians
accounting for interactions with the increasing number of neighbors.
(For a single band it was first noticed in  \cite{semn} ).  
In the two band model  \cite{seven}  
 the first order term in $\mid A\mid$ from (\ref{eq:a6}) is $(\langle
S_z\rangle =3/2)$:
\begin{equation}
\hbar\omega({\bf k})=|A|(3-c_x-c_y-c_z)D(x)/3
\label{eq:a9}
\end{equation}
and $D(x)\equiv D(E(x))$ is given by the integral:
\begin{eqnarray}
\int\frac{d^3{\bf p}}{(2\pi)^3}\left[\, 
\sum_{(+,-)}\theta(E-\varepsilon_i({\bf p}))
\left\{
1\pm\frac{2c_x-c_y-c_z}{2 R({\bf p})}\right\}
\right ],
\nonumber
\end{eqnarray}
(here $E$ is in units of $|A|, p_i\equiv a p_i$).  
Quantum fluctuations may change  the {$\bf k$} -- dependence 
in Eq. (\ref{eq:a9}).

To calculate the conductivity, $\sigma(\omega, x)$, 
we determine first the velocity operator,
$\hat{{\bf v}}=\hat{{\bf \dot{r}}}$ (see \cite{vose}):
\begin{equation}
\hat{\bf v}({\bf k})=\frac{1}{\hbar}\frac{\partial\varepsilon_l({\bf k})}{\partial
{\bf k}}+\frac{i}{\hbar}[\varepsilon_l({\bf k})-\varepsilon_{l'}({\bf k})]\,\langle
l{\bf k}|\hat{{\boldmath \Omega}}|l'{\bf k}\rangle.
\label{eq:a10}
\end{equation}
The off-diagonal operator $\hat{{\boldmath \Omega}}$ is
defined by the relation:
\begin{equation}
\langle l{\bf k}|{\boldmath \Omega} |l'{\bf k}\rangle =i\int
u_{\bf k}^{\ast l'}({\bf r})\frac{\partial
u_{\bf k}^l}{\partial {\bf k}} d^3{\bf r}
\label{eq:a11}
\end{equation}
and $u_{\bf k}^l({\bf r})$, the periodic Bloch functions on the basis
 \cite{seven}  are:
\begin{eqnarray}
u_{\bf k}^l({\bf r})&=&\frac{1}{\sqrt{N}}\sum_n\exp
[i{\bf k}({\bf a}n-{\bf r})]\times\nonumber \\
&& \times ~\left\{\alpha_{\bf k}^l\phi_1({\bf r}-n{\bf
a})+\beta_{\bf k}^l\phi_2({\bf r}-n{\bf a})\right\}.
\label{eq:a12}
\end{eqnarray}
With the one-site integrals only in (\ref{eq:a11}) and 
Eqs. (\ref{eq:a8}):
\begin{equation}
\langle l{\bf k}|\hat{{\boldmath \Omega}}|l'{\bf k}\rangle
=i\frac{a}{\hbar}\frac{\sqrt{3}}{4}\frac{(-\sin
k_x)(c_y-c_z)}{|t_{12}|^2}.
\label{eq:a14}
\end{equation}
Eqs. (\ref{eq:a10}, \ref{eq:a14}) produce transitions from {\it
occupied} parts of the $\varepsilon_+({\bf p})$-band into {\it empty} states
in the $\varepsilon_-({\bf p})$-band.

With all the above, we arrive to the Drude (intraband)
contribution which in the clean limit is:
\begin{eqnarray}
\sigma_{Drude}(\omega, x)&=&2\pi\frac{e^2 |A|}{3a\hbar^2}\delta(\omega)
I_{Dr}(x), \\
I_{Dr}(x)&=&\frac{1}{2(2\pi)^3}\sum_l\int dS_{\bf p}^l|{\boldmath
\nabla}_{\bf p}\varepsilon({\bf p})|,
\label{eq:a15}
\end{eqnarray}
(the integral in $I_{Dr}(x)$ is over the Fermi surfaces (FS)).\\
The ``optical'' (interband) contribution is 
\begin{eqnarray}
\sigma_{opt}&(&\omega,x)=\frac{3\pi e^2}{a\hbar}\frac{1}{\tilde{\omega}_0^3}
\int\frac{d^3{\bf p}}{(2\pi)^3}\sin^2p_x(c_y-c_x)^2 
 \times \nonumber \\ 
&& \times ~n(\varepsilon_+({\bf p}))[\,1-n(\varepsilon_-({\bf p}))]  
\cdot \delta\left(
\tilde{\omega}-2 R({\bf p})\right),
\label{eq:a16}
\end{eqnarray}
where $|A|\tilde{\omega}=\hbar\omega$ (Eqs. (\ref{eq:a15}) and 
(\ref{eq:a16}) agree with the results of  \cite{devj}).
For the low temperature spectral
weights $N_{eff}={{2m}\over{\pi e^{2}}} a^3 
\int\limits_{0}^{\infty}\sigma(\omega) d\omega$, 
one obtains for the Drude and the interband contributions,
respectively:
\begin{eqnarray}
N_{eff}^{Drude}=\frac{ma^2}{3\hbar^2}|A|I_{Dr}(x),
\label{eq:adrude} 
 ~~N_{eff}^{opt}=\frac{ma^2}{\hbar^2}|A|\frac{3}{4}I_{opt}(x),
\label{eq:aopt}
\end{eqnarray}
(with $I_{opt}(x)$ directly obtained from (\ref{eq:a16})).
In Fig. 1 we plotted our results for the Fermi level, 
$E_{F}(x) = |A|E(x), 
\tilde{\nu}(x)={\nu}(x)|A|, 
D(x)$ and $I_{Dr}(x)$.
In Fig. 2 the Fermi surface at x=0.3 is shown (the shaded area
shows schematically the concentration range for a percolative
regime). 

{\it Fermi Liquid and the Experimental Data}.-- We choose to determine the
band parameter, $|A|$, from measurements of the spin stiffness
coefficient: $\hbar\omega({\bf k})=({\bf
k}a)^2|A| D(x)/6$. As the long-wave
characteristics, $D(x, T=0)$ is less sensitive to a samples'
disorder.  
We use $a\simeq 3.86\AA$, $D(x\simeq 0.3)\simeq 0.45$
and data in  \cite{chet}  (Table 1).
The results for $|A|$ 
show that the bandwidths for different materials, $W$=6$|A|$, 
do not vary significantly ($0.7\,-\,1.0$ eV) as if changes in 
the Mn-O-bond angle
in different materials are {\it not} of importance.
The differences in ionic radii regulate mainly 
 the local disorder as seen from
data on resistivity (see below).  
According to \cite{moudden}, $D(x)$ first increases with
$x$ above $x_{cr}$ and saturates at $x \sim 0.3$. The
initial increase of $D(x)$ \cite{moudden} 
merely reflects the fact
that above the percolation threshold the percentage 
of the FM-phase increases (magnetic moment increases).
In the FM-phase further increase in $x$ leads to {\it decrease}
of the magnetization and $D(x)$ in Fig. 1 decreases.
This decrease is beyond the accuracy of data \cite{moudden},
and the crossover into FM state is seen as a saturation 
of $D(x)$ at $x=0.28, 0.3$ \cite{moudden}. 

\begin{figure}
\centerline
{\psfig{figure=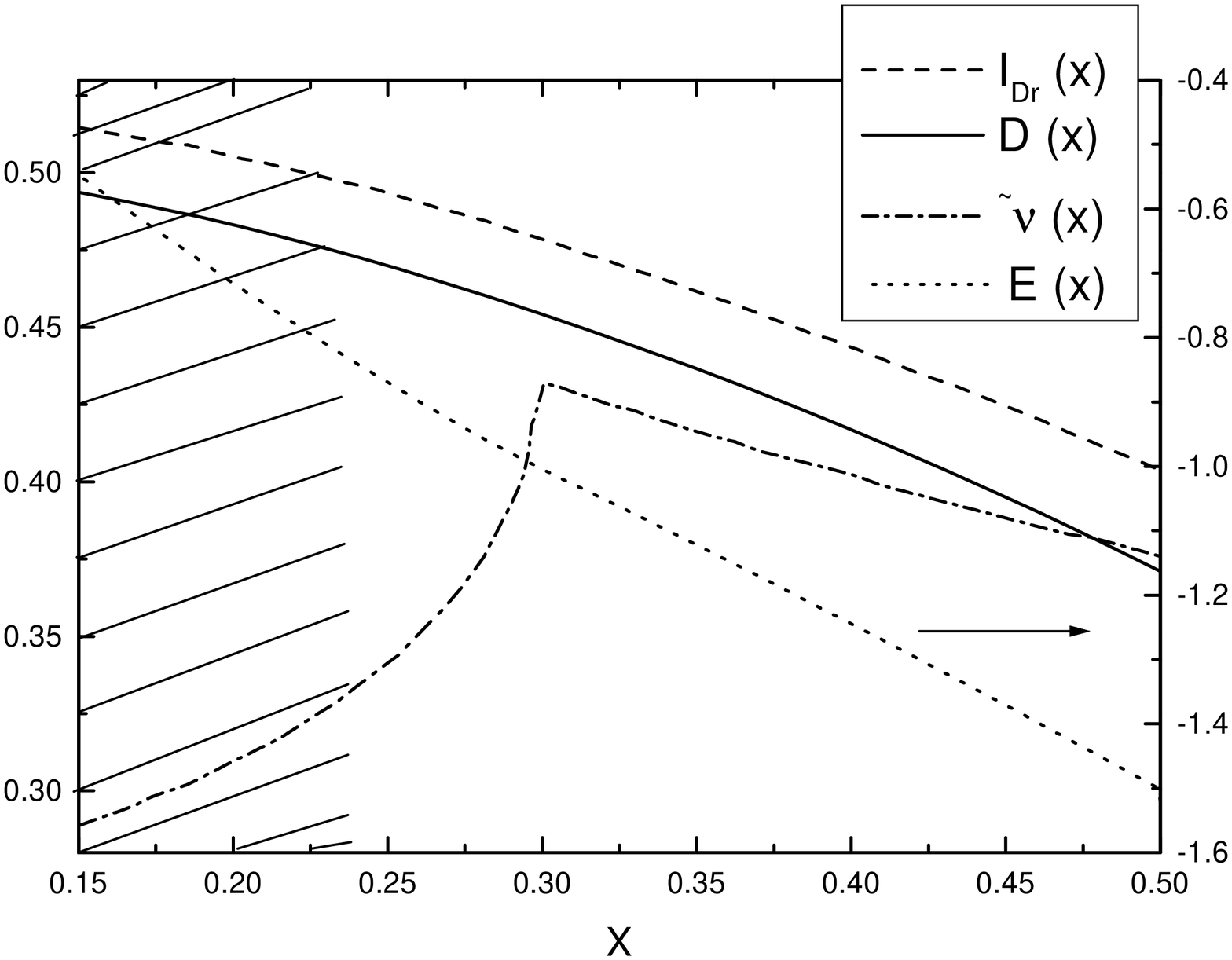,height=6cm,width=7.5cm,angle=0}}
\caption{The Fermi level, E(x), DOS, $\tilde{\nu}(x)$, the spin
stiffness coefficient D(x) and the Drude conductivity, 
$I_{Dr} (x)$, plotted as a function of concentration, x, 
for the spectrum Eq. (2). The shaded area shows the percolative
regime where Eqs. (9-18) are not applicable.}
\end{figure}

Another important parameter, the Sommerfeld coefficient,
$\gamma$, for our spectrum (\ref{eq:a2}) is:
\begin{eqnarray}
\gamma =\pi^2\tilde{\nu}(x)/3|A|.
\label{eq:a18}
\end{eqnarray}
Eq. (\ref{eq:a18}) gives 
$\gamma\sim 6.2$ mJ/mole$\cdot K^2$ ($\tilde{\nu}(x)\sim 0.45$ at
$x\simeq 0.3$). According to  \cite{dvad} , 
$\gamma\simeq 3.5$ mJ/mole$\cdot K^2$
for La$_{0.7}$Sr$_{0.3}$MnO$_3$. 
The low-T heat capacity of a few other manganites with compositions
$x\simeq 0.3$ has been measured with $\gamma$ in the 3-8 mJ/mole$\cdot K^2$
range \cite{one}.  

For most of experimental
results it is common (e.g.  \cite{one,dvad}\,) to ``separate'' the
$T^{3/2}$ magnon term in the specific heat.  
Here lies an interesting catch.  Namely, close to x=0.3 in
addition to that of magnons, 
there is the {\it band}  $T^{3/2}$-contribution into the
specific heat.  Indeed, the Fermi surface in Fig. 2, shows a
formation of the ``neck'' at the zone boundary very close to $x=0.3$,
and, hence, this is the point of the ``2.5''-Lifshitz transition!  With
this in mind, one should recognize a rather good agreement between 
(\ref{eq:a18}) and the low-T experimental data on the heat capacity.
As it is easy to see, $\tilde\nu(x)$ in Fig. 1 varies significantly 
near x=0.3.

Substitution of divalent atoms at such concentrations
inevitably leads to intrinsic disorder.  It is argued in 
 \cite{nine}  that
this disorder, seen in La$_{1-x}$Sr$_x$MnO$_3$ even at T=10 K at
$x<0.35$ is mainly due to the local JT-distortions 
and bears a quasistatic character at $x>0.17$.  
Part of carriers may be trapped into the JT polarons, 
forming the microdomains of an insulating phase.
The disorder is 
a factor, which may affect our interpretation.  In view that
resistivities for samples with nominally the same concentration may
differ in the order of magnitude, it is essential to evaluate related
spatial and energy scales. We substitute in
Eq.(\ref{eq:a15}) 
$\pi\delta(\omega)\rightarrow{\tau}/{(1+(\omega\tau)^2)}$
to obtain the effective $\hbar /\tau$ due to residual
resistivities.  This gives  $\hbar/\tau |A|\sim 0.5$ for
materials with $\rho\simeq 300(\mu\Omega -cm)$, but it is only $\sim
2\cdot 10^{-2}$ for the film samples of 
La$_{0.7}$Sr$_{0.3}$MnO$_3$\cite{six}.  
With the average velocity of an electron on the FS in our model
$\langle {\bf v}^2\rangle^{1/2}=\left( |A|a/\hbar \right )
\left(2 I_{Dr}(x)/
\tilde{\nu}(x)\right)^{1/2}$, 
the mean free path, $l$, 
is typically of the order of $l\sim 3 a$ for most
samples, while $l\sim 80 a$ is reached
in the best LSMO-samples  \cite{six} .  
It remains to be seen whether sample's quality may be further
improved.

\begin{figure}
\centerline
{\psfig{figure=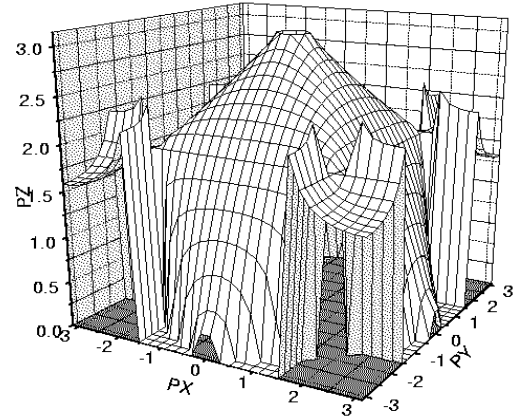,height=10cm,width=8.5cm,angle=0}}
\vspace{-3cm}
\caption{The FS at $x\approx 0.3$.``Necks'' forming at the
zone boundary are responsible for singularity in ${\tilde{\nu}}(x)$
in Fig. 1.}
\end{figure}

The $T^2$ -- term in resistivity due to the electron interactions comes
from 
$\hbar/{{\tau}_{ee}}^{tr} = {\lambda}^{\prime}\left(\hbar/{\tau}_{ee}
\right)$, where ${\tau}_{ee}$ is the total quasiparticle 
relaxation time and ${\lambda}^{\prime}<1$ gives the fraction 
of Umklapp processes.
We use \cite{rashba}:
\begin{eqnarray}
\hbar /\tau_{ee}\simeq \lambda{\pi}^3{\nu}(x)T^2 .
\label{eq:a21}
\end{eqnarray}
In (\ref{eq:a21}) $\lambda$ is a value of the interaction strength
in terms of $E_F$. Using \cite{six,odin}, one obtains for LSMO:
${\lambda}{\lambda}^{\prime}\simeq 0.3$, typical of good metals.
As for the two other materials in \cite{six}, the $T$-variations
of resistivity scale in the magnitude with their residual resistivity
and are caused by defects \cite{mahan}.

$T$- dependence in the optical conductivity, $\sigma(\omega)$,
attracted recently much attention  \cite{six,shes,dvad} 
as a manifestation of changes in the conductivity mechanism at
elevated temperatures. We discuss only a few results pertinent
to the low $T$ band mechanisms. First note, that the
temperature dependence in $\sigma(\omega)$ at
$T<100$ K for $\omega$ around 1eV is most pronounced 
{\it below} 1eV (see Fig. 2 in \cite{six} ).  
This agrees well with our estimates for the bandwidths,
$W \leq 1$eV.  As for a quantitative analysis, there
are problems of an experimental character.  
Assuming $N_{eff}(\omega)$  \cite{six}  would give our
$N_{eff}$'s in Eq. (\ref{eq:adrude}) 
at $\omega\simeq 1$eV and that both the
Drude and optical contributions are approximately equal \cite{devj}, 
we obtain
$N_{eff}\sim 0.25$ which is reasonably close to the values 
in  \cite{six} ,
lesser than $N_{eff}$ for the single crystal data,
La$_{0.67}$Ca$_{0.33}$MnO$_3$  \cite{shes} , and a factor ten bigger than
$N_{eff}$ for La$_{0.7}$Sr$_{0.3}$MnO$_3$ 
in  \cite{pjat} .  We believe that such
a difference originates from poor data for the optical $\sigma
(\omega)$ in the ``Drude-tail'' range.

Finally, the optical gap for pure LaMnO$_3$ was 
identified  \cite{pjat}  at
$\Delta\simeq 1.2$eV.  A rough estimate from the band insulator picture
 \cite{eight}  gives $g\sim 0.6$eV, i.e. the JT-coupling is rather 
strong ($|A|\sim0.16$eV).

We conclude the discussion by a few comments regarding the spin
wave spectrum.  In La$_{0.7}$Pb$_{0.3}$MnO$_3$ 
 \cite{dve}  the spectrum fits
well (\ref{eq:a9}). Eq.(\ref{eq:a9}) follows from (\ref{eq:a6}) 
at $A \ll J_H\langle S_z\rangle$, with quantum corrections neglected.
Meanwhile, strong deviations from (\ref{eq:a9}) have been
observed at $\xi\,>\,0.25$ along the (0,0,$\xi$)-direction 
in Pr$_{0.63}$Sr$_{0.37}$MnO$_3$  \cite{tri}. The spin stiffness
changes only slightly from material to material \cite{chet},  
and other low
temperature characteristics including electron 
interactions in (\ref{eq:a21}),
also seem not to vary much. 
Unlike \cite{chet}, we suggest with \cite{tri} that 
differences in low-$T$ spin dynamics for
two materials do not correlate with their behavior at higher temperatures,
but with a tendency to a low temperature charge ordering at
$x$ close to 0.5.

In conclusion, we investigated low temperature properties of different
manganites with hole concentration close to $x\simeq 0.3$. Not only the
tight-binding model \cite{seven} demonstrates applicability of the Fermi
liquid approach at $T\rightarrow 0$, but also it agrees 
with most experimental data, suggesting thus the unifying approach
to ferromagnetic manganites.  
Bandwidths for different
materials change only slightly.  Concentration $x\simeq 0.3$ is
remarkable for the 2.5 -- topological transition. 
Samples quality remains a major obstacle to
further elaboration of the low temperature properties.
In some materials conductivity falls into the mobility
edge regime.

L.P.G. and V.Z.K. gratefully acknowledge H.D. Drew,
T. Egami, J. Lynn, S. von Molnar for
numerous stimulating discussions. L. P. G. thanks 
S. Watts for essential help with some calculations. 
The work was supported (L.P.G. and M.O.D.) by the National High Magnetic
Field Laboratory through NSF Cooperative Agreement \# DMR-9527035 and the
State of Florida.

\end{document}